\documentclass[%
 reprint,
 amsmath,amssymb,
 aps,
]{revtex4-2}

\usepackage{graphicx}
\usepackage{dcolumn}
\usepackage{bm}
\usepackage{booktabs}
\usepackage{array}
\usepackage{xcolor}
\usepackage[utf8]{inputenc}
\usepackage[T1]{fontenc}
\usepackage[english]{babel}
\usepackage{multirow}

\usepackage{titlesec}

\titlespacing*{\section}
{0pt}   
{4.0ex} 
{4.0ex} 

\titlespacing*{\subsection}
{0pt}   
{3.0ex} 
{3.0ex} 

\begin{document}

\preprint{APS/123-QED}

\title{\textbf{Anomalous magnetotransport in a non-collinear correlated kagome ferromagnet MgMn$_6$Sn$_6$}}

\author{Kakan Deb$^{1}$, Sourav Kanthal$^{1}$,  Jyotirmoy Sau$^{1}$, Chandra Shekhar$^{2}$, Manoranjan Kumar$^{1}$, Matthias Gutmann$^{3}$, Jhuma Sannigrahi$^{4,*}$ and Nitesh Kumar$^{1,*}$}
\affiliation{$^{1}$
S. N. Bose National Centre for Basic Sciences, 
Salt Lake, Kolkata 700106, India}
\affiliation{$^{2}$Max Planck Institute for Chemical Physics of Solids, 01187 Dresden, Germany}
\affiliation{$^{3}$ISIS Neutron and Muon Source, Science and Technology Facilities Council, 
Rutherford Appleton Laboratory, Chilton, Didcot OX11 0QX, United Kingdom}
\affiliation{$^{4}$School of Physical Sciences, Indian Institute of Technology Goa, 
Farmagudi, Goa 403401, India}

\begin{abstract}
Magnetic kagome metals provide a fertile platform for exploring unusual magnetotransport phenomena arising from the intricate interplay between electronic topology, electron correlations, and magnetic order. MgMn$_6$Sn$_6$ is a room-temperature kagome ferromagnet with strong in-plane magnetic anisotropy. Here, we report a combined study of single-crystal neutron diffraction (SCND) and magnetotransport properties of MgMn$_6$Sn$_6$ backed by first principles calculations. Our SCND measurements reveal a non-collinear arrangement of Mn magnetic moments within the basal plane of the kagome bilayer. Hall conductivity yields a substantial intrinsic contribution of approximately $0.29\,e^2/h$ per kagome layer, which is nearly isotropic with respect to the field orientation. At low temperatures, the anomalous Hall conductivity develops a pronounced anisotropic extrinsic component, highlighting the directional sensitivity of scattering processes. The significantly large value of the Sommerfeld coefficient in the absence of $f$-electrons underscores an enhanced electron correlation. Therefore, the non-collinear kagome ferromagnet MgMn$_6$Sn$_6$ is a good candidate to study the effect of correlation on the magneto-transport properties. 
\end{abstract}

\maketitle 

\section{\label{sec:level1}INTRODUCTION} 
The investigation of correlated topological electronic states, arising from the interplay of lattice geometry, nontrivial electronic band topology, and electron-electron interactions, has become a key focus in modern condensed matter physics \cite{tokura2017emergent,yin2018giant,paschen2021quantum,morosan2012strongly}. Among the various material platforms, compounds based on the kagome lattice have attracted considerable attention due to their unique geometry of corner-sharing triangles. This lattice geometry naturally leads to an electronic structure featuring nearly flat bands, symmetry-protected Dirac dispersions, and van Hove singularities in the density of states \cite{wang2023quantum,di2026kagome,li2018realization,kang2020dirac,hu2022tunable}. When the Fermi energy ($E_F$) lies close to these singular features, electronic correlations are often strongly enhanced, which can drive collective phenomena such as charge density wave order, spin density wave order, unconventional superconductivity, etc \cite{neupert2022charge,teng2022discovery,tazai2022mechanism,kang2023charge,park2025spin}. Moreover, presence of magnetic order and spin-orbit coupling (SOC) produces significant Berry curvature, which in turn leads to various exotic quantum transport responses, including large anomalous Hall effect (AHE) and Nernst effect. \cite{karplus1954hall,smit1955spontaneous,smit1958spontaneous,nagaosa2010anomalous,liu2018giant,li2017anomalous,asaba2021colossal}.

Although kagome lattices support a variety of frustrated magnetic states, only a few systems, such as Co$_3$Sn$_2$S$_2$, Fe$_3$Sn$_2$, Fe$_3$Sn, and LnTi$_3$Bi$_4$ (Ln = Nd, Sm, Eu), exhibit ferromagnetism \cite{liu2018giant,ye2018massive,PhysRevB.94.075135,du2022topological,belbase2023large,ortiz2023evolution}. Among these, Co$_3$Sn$_2$S$_2$ shows a large intrinsic anomalous Hall conductivity (AHC) of $\sim1130~(\Omega\cdot\mathrm{cm})^{-1}$ arising from SOC-gapped nodal-line crossings and Weyl nodes located close to $E_F$ \cite{liu2018giant}, while Fe$_3$Sn$_2$ hosts massive Dirac fermions and exhibits both a sizable intrinsic anomalous Hall response of $\sim302~(\Omega\cdot\mathrm{cm})^{-1}$and a large topological Hall effect \cite{ye2018massive,PhysRevB.94.075135,du2022topological}.

\begin{figure*}[t]
    \centering
    \includegraphics[width=1\textwidth]{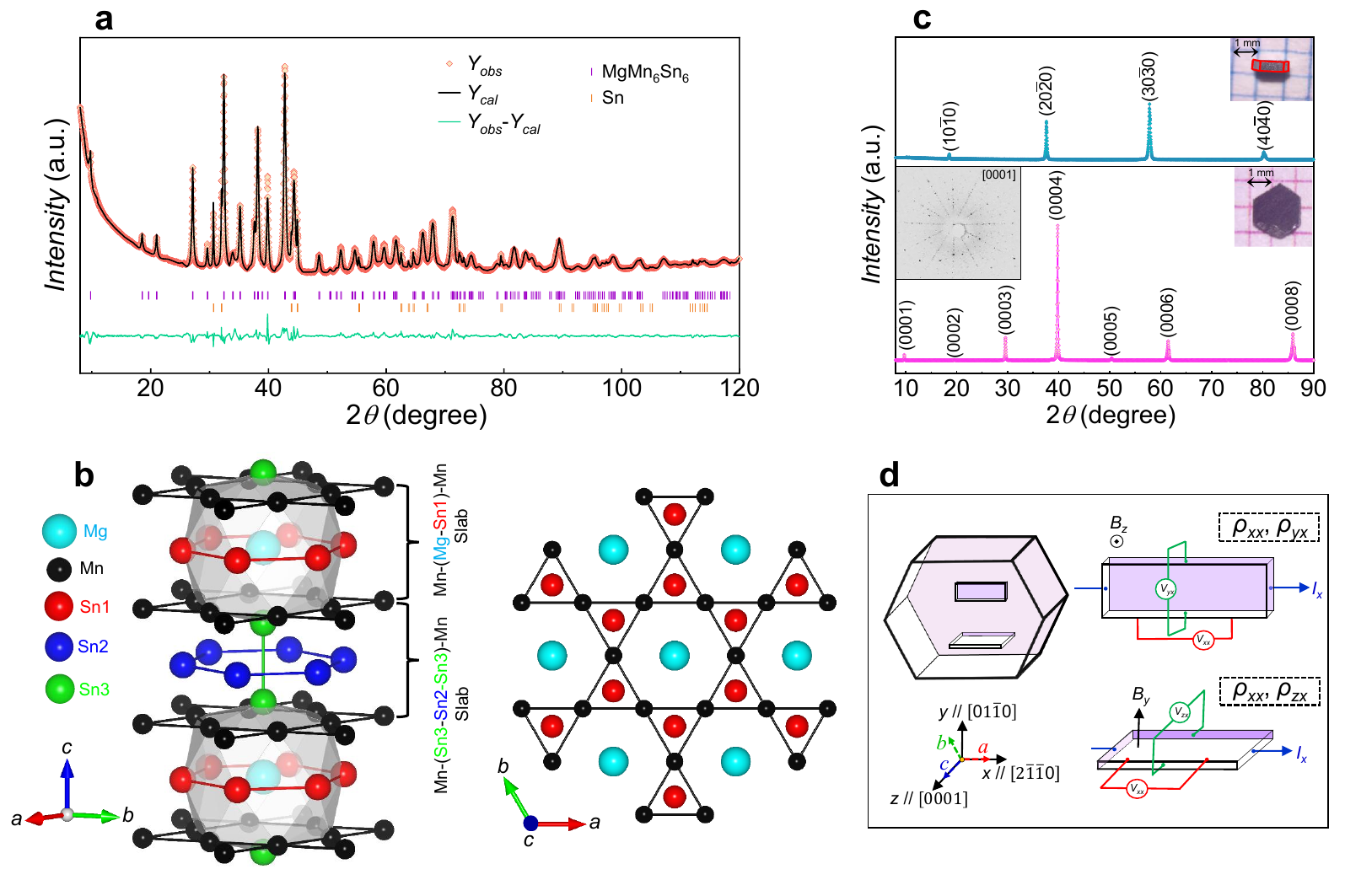}
    \caption{\textbf{a}, Powder XRD pattern of $\mathrm{MgMn_6Sn_6}$ along with the Rietveld refinement. \textbf{b}, Crystal structure of $\mathrm{MgMn_6Sn_6}$ viewed from the side and top. \textbf{c}, In-plane and out-of-plane X-ray diffraction patterns of $\mathrm{MgMn_6Sn_6}$ single crystals. The right insets show photographs of typical single crystals, while the bottom-left inset shows a Laue diffraction photograph taken along the $[0001]$ direction of an aligned sample. \textbf{d}, Schematic diagram of the crystal shape, orientation, and the associated orthogonal axes, along with magnetotransport measurement configurations: $\rho_{yx}$ (\( I \parallel x \), \( B \parallel z \)) and $\rho_{zx}$ (\( I \parallel x \), \( B \parallel y \)).}
   
    \label{fig:fig1}
\end{figure*}

\textit{R}Mn$_6$Sn$_6$ family (where \textit{R} denotes a rare-earth element), consisting of Mn kagome bilayers, has emerged as an important platform to study the coupling between magnetism and band topology \cite{PhysRevLett.126.246602,PhysRevB.108.045132,riberolles2024new}. This material system hosts a variety of magnetic ground states, making it particularly suitable for disentangling the roles of spin, orbital, and topological degrees of freedom in transport and spectroscopic responses. In compounds where \textit{R} is a magnetic rare-earth ion (\textit{R} = Gd-Ho), strong antiferromagnetic (AFM) coupling between the \textit{R} and Mn sublattices stabilizes a collinear ferrimagnetic (FIM) order at low temperatures, with distinct magnetic anisotropies: easy-plane for Gd, easy-axis for Tb, and easy-cone for Dy and Ho \cite{asaba2020anomalous, yin2020quantum,jones2024origin,gao2021anomalous}. For \textit{R} = Er and Tm (weakly-magnetic), the \textit{R}-Mn magnetic coupling is significantly weaker than the interlayer AFM Mn-Mn interaction, resulting in an in-plane FIM (Er) and AFM (Tm) order at low temperatures \cite{PhysRevMaterials.8.094411,PhysRevB.106.125107}. Meanwhile, compounds with nonmagnetic rare-earth ions (\textit{R} = Sc, Y, Lu) exhibit complex magnetism characterized by spiral AFM ordering, driven by competing interlayer magnetic couplings between nearest- and next-nearest-neighbor Mn kagome layers \cite{zhang2022magnetic,ghimire2020competing,hwz4-z9mm}.

A large intrinsic AHE has been observed across various \textit{R}Mn$_6$Sn$_6$ compounds (\textit{R} = Gd-Er), originating from the Berry curvature associated with Chern gaps near $E_F$, as found in the quantum-limit Chern magnet TbMn$_6$Sn$_6$ \cite{yin2020quantum,PhysRevLett.126.246602}. In addition to the AHE, a topological Hall effec (THE) has also been reported in \textit{R}Mn$_6$Sn$_6$ compounds, which is attributed to a finite scalar spin chirality (SSC) arising either from chiral spin textures or from thermally driven spin fluctuations \cite{PhysRevMaterials.8.094411,PhysRevB.106.125107,zhang2022magnetic,ghimire2020competing,hwz4-z9mm}.

Replacing the rare-earth element (\textit{R}) with alkali or alkaline-earth metals (e.g., Mg and Li) induces ferromagnetic (FM) ordering, a feature notably absent in rare-earth-based \textit{R}Mn$_6$Sn$_6$ compounds (except for non-magnetic \textit{R} = Yb) \cite{mazet1998magnetic,PhysRevB.103.144410,mazet1999study}. In this work, we focus on MgMn$_6$Sn$_6$, a rare example of a room-temperature kagome ferromagnet. Although previous studies have suggested soft, easy-plane ferromagnetism based on bulk magnetization measurements, a microscopic determination of its magnetic structure has remained elusive \cite{song2024magnetocaloric,song2024critical}.

In the present study, we resolve the magnetic structure of MgMn$_6$Sn$_6$ using single-crystal neutron diffraction (SCND) and analyze the magnetic symmetry group. We further investigate the electronic transport properties through Hall-effect measurements, supported by density functional theory (DFT) calculations. Moreover, MgMn$_6$Sn$_6$ exhibits significantly large electronic heat capacity, indicating enhanced electron correlation effects in this Mn-based layered kagome system.
\section{\label{sec:level2} RESULTS AND DISCUSSION} 

\subsection{Crystal Structure}

MgMn$_6$Sn$_6$ crystallizes in the hexagonal HfFe$_6$Ge$_6$-type structure with space group \textit{P}6/\textit{mmm} \cite{mazet2002r}. X-ray diffraction (XRD) pattern obtained on the powdered single crystals at room temperature is shown in Fig.~1\textbf{a} along with the generated Rietveld refined pattern. The refined cell parameters are \(a = 5.5199(2)\,\text{\AA}\) and \(c = 9.0429(3)\,\text{\AA}\). The structural parameters obtained from the refinement are listed in Tables~I. Small amount of Sn impurity phase (less than 5\%) largely present on the crystal surface as excess flux, has been included in the Rietveld refinement. 

\begin{table}[t]
\centering
{
\begin{tabular}{@{}lcccccc@{}}
\specialrule{0.4pt}{0pt}{4pt}
Atom    & Wyckoff site & $x$     & $y$     & $z$        & Occ.     & $U_{\text{iso}}$ (\AA$^2$)\\ \specialrule{0.4pt}{2pt}{4pt}
Mg      & 1a               & 0       & 0       & 0          & 1          & 0.038(5) \\
Mn      & 6i               & 0.5     & 0.5     & 0.7507(3)  & 1          & 0.0010(9) \\
Sn1     & 2c               & 0.3333  & 0.6667  & 0          & 1          & 0.0159(10) \\
Sn2     & 2d               & 0.6667  & 0.3333  & 0.5        & 1          & 0.0108(10) \\
Sn3     & 2e               & 0       & 0       & 0.3258(2)  & 1          & 0.0144(8) \\ 
\specialrule{0.4pt}{3pt}{2pt}
\end{tabular}
 }
\caption{Crystallographic data determined for $\text{MgMn}_6\text{Sn}_6$ in the \( P6/mmm \) hexagonal space group symmetry.}
\label{tab:atomic_coordinates}
\end{table}

The crystal structure of this compound can be described as a stuffed variant of the CoSn-type structure \cite{venturini2006filling}, in which adjacent Mn-based kagome layers are separated by alternating nonmagnetic Sn$_3$-Sn$_2$-Sn$_3$ trilayers and mixed Mg-Sn$_1$ layers (Fig.~1\textbf{b}). The intralayer nearest Mn-Mn distance is $2.759$~\AA{}, whereas the interlayer Mn-Mn separations across the Sn trilayers and the Mg-Sn mixed layers are $4.534$~\AA\ and $4.509$~\AA, respectively. The Sn$_1$ and Sn$_2$ atoms each form separate honeycomb layers, with Sn$_2$ located directly beneath Sn$_1$ along the $c$ axis, making it invisible in the top view (right panel of Fig.~1\textbf{b}). Mg atoms occupy the centers of the Sn$_1$ hexagonal units, and the chemical pressure exerted by Mg displaces the Sn$_3$ atoms by $\sim 0.713$~\AA{} from the centers of the Mn kagome layers. This displacement creates large hexagonal void spaces, and also results in pristine Mn-based kagome planes, which are absent in the CoSn structure. This makes MgMn$_6$Sn$_6$ an ideal platform for exploring kagome physics. In general, for \textit{R}Mn$_6$Sn$_6$ compounds, alternating Mn-Sn$_3$-Sn$_2$-Sn$_3$-Mn and Mn-(\textit{R}, Sn$_1$)-Mn slabs construct the three-dimensional crystal framework. The Mn-Mn interlayer magnetic coupling depends on the choice of \textit{R} atoms. In MgMn$_6$Sn$_6$, the lower valence of Mg compared to rare-earth ions favors FM interlayer coupling between the Mn kagome layers \cite{mazet2002r}, similar to that observed in LiMn$_6$Sn$_6$ \cite{PhysRevB.103.144410}.

\subsection{ Magnetism and magnetic structure }%
Temperature-dependent magnetization (\(M\)), measured under an applied magnetic field of 50 Oe along the \(x\) and \(z\) axes in the field-cooled warming (FCW) mode (Fig.~2\textbf{a}), shows a paramagnetic-to ferromagnetic transition with a Curie temperature \( T_\mathrm{C} \approx 300~\mathrm{K} \), in agreement with previous reports \cite{song2024magnetocaloric,mazet1999study}. Below \( T_\mathrm{C} \), the magnetization for \( B \parallel x \) is significantly larger than that for \( B \parallel z \), indicating that the \textit{ab}-plane is the magnetic easy plane, while the \textit{c}-axis corresponds to the hard axis. Figs.~2\textbf{b} and 2\textbf{c} display the isothermal magnetization curves measured at various temperatures for \( B \parallel x \)  and \( B \parallel z \), respectively. Below \( T_{\mathrm{C}} \), for both the orientations \(M\) increases linearly at low fields and saturates beyond a characteristic saturation field (\( B_{\mathrm{s}} \)). At \( T = 2\,\text{K} \), a pronounced anisotropy is evident from the markedly different \textbf{\( B_{\mathrm{s}} \)}, with \( B_{\mathrm{s}}^{x} = 0.25\,\text{T} \) for \( B \parallel x \) and \( B_{\mathrm{s}}^{z} = 2.5\,\text{T} \) for \( B \parallel z \), underscoring a strong magnetic anisotropy favoring in-plane alignment of magnetic moments. The saturated moment at 2\,K, \( M_{\mathrm{S}} \approx 2.15\,\mu_{\mathrm{B}}/\text{Mn} \), is in good agreement with previously reported values \cite{mazet2002r,song2024critical}. To further investigate the magnetic anisotropy, the magnetic  anisotropy energy density $K_U$ was obtained using 

\begin{equation}
K_U = \mu_0 \int_{0}^{M_s} [B_x(M) - B_z(M)]\, dM
\end{equation}
where $M_s$ is the saturation magnetization, and $B_x$ and $B_z$ correspond to the fields applied along the $x$ and $z$ directions, respectively (Inset of Fig.~2\textbf{d}). The temperature dependence of \( K_U \) shows a monotonic increase upon cooling, reaching a maximum value of \( 5.29 \times 10^{5}\,\mathrm{J\,m^{-3}} \) at 2~K and decreasing to \( 0.27 \times 10^{5}\,\mathrm{J\,m^{-3}} \) at room temperature \cite{pal2025ferromagnetic}. These findings suggest that MgMn$_6$Sn$_6$ exhibits a pronounced magnetocrystalline anisotropy (MCA) persisting to near room temperature, which is advantageous for practical applications.

\begin{figure}[t]
\centering
\includegraphics[width=\linewidth]{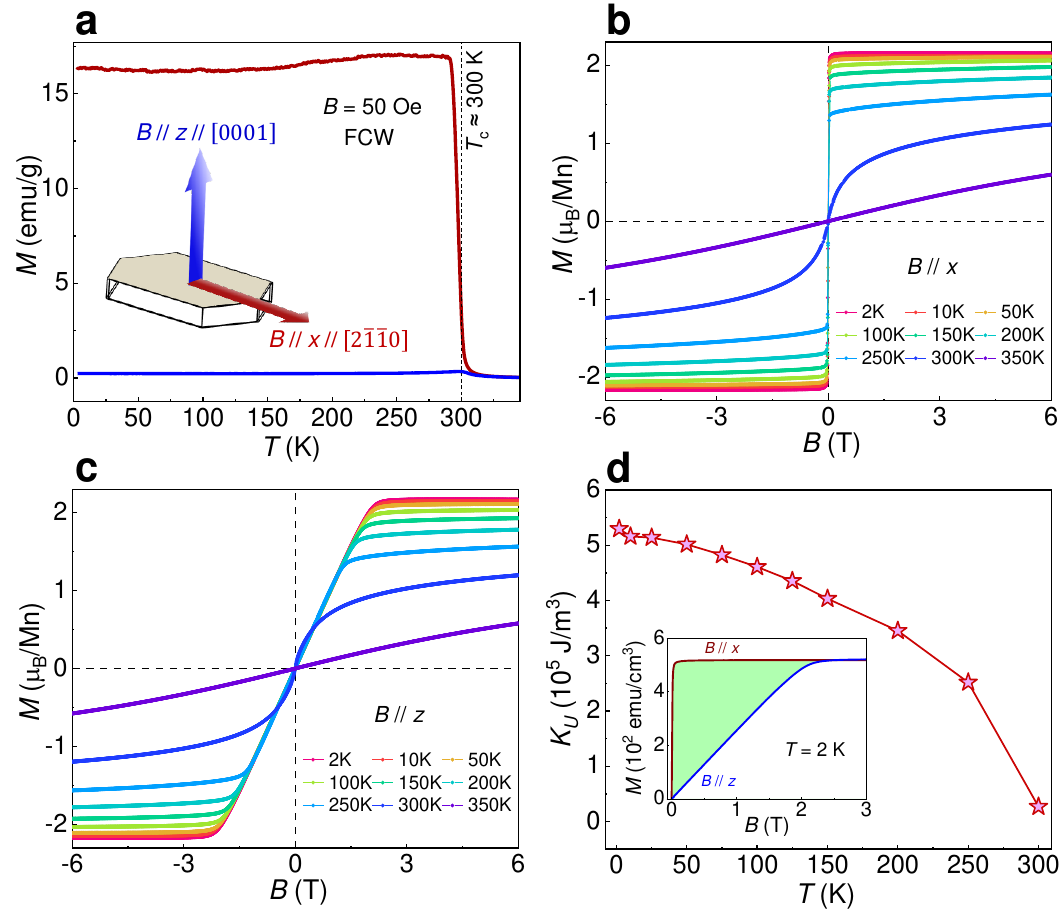}
\caption{\textbf{a}, Magnetization measured under an external magnetic field $B = 50~\mathrm{Oe}$ applied along the $x$ and $z$ directions, following the FCW protocol. \textbf{b}, \textbf{c}, Magnetization as a function of magnetic field, with $B \parallel x$ (\textbf{b}) and $B \parallel z$ (\textbf{c}), recorded at selected temperatures between 2-350~K. \textbf{d}, Temperature dependence of the magnetic anisotropy energy density $K_U$. The inset shows the $M$-$B$ curves measured at 2~K, and the shaded region corresponds to the magnetic anisotropy energy density, $K_U = \mu_0 \int_{0}^{M_s} [B_x(M) - B_z(M)]\, dM$.}
\label{fig:fig2}
\vspace{-0.6cm}
\end{figure}

\begin{figure*}[t]
\centering
\includegraphics[width=1\textwidth]{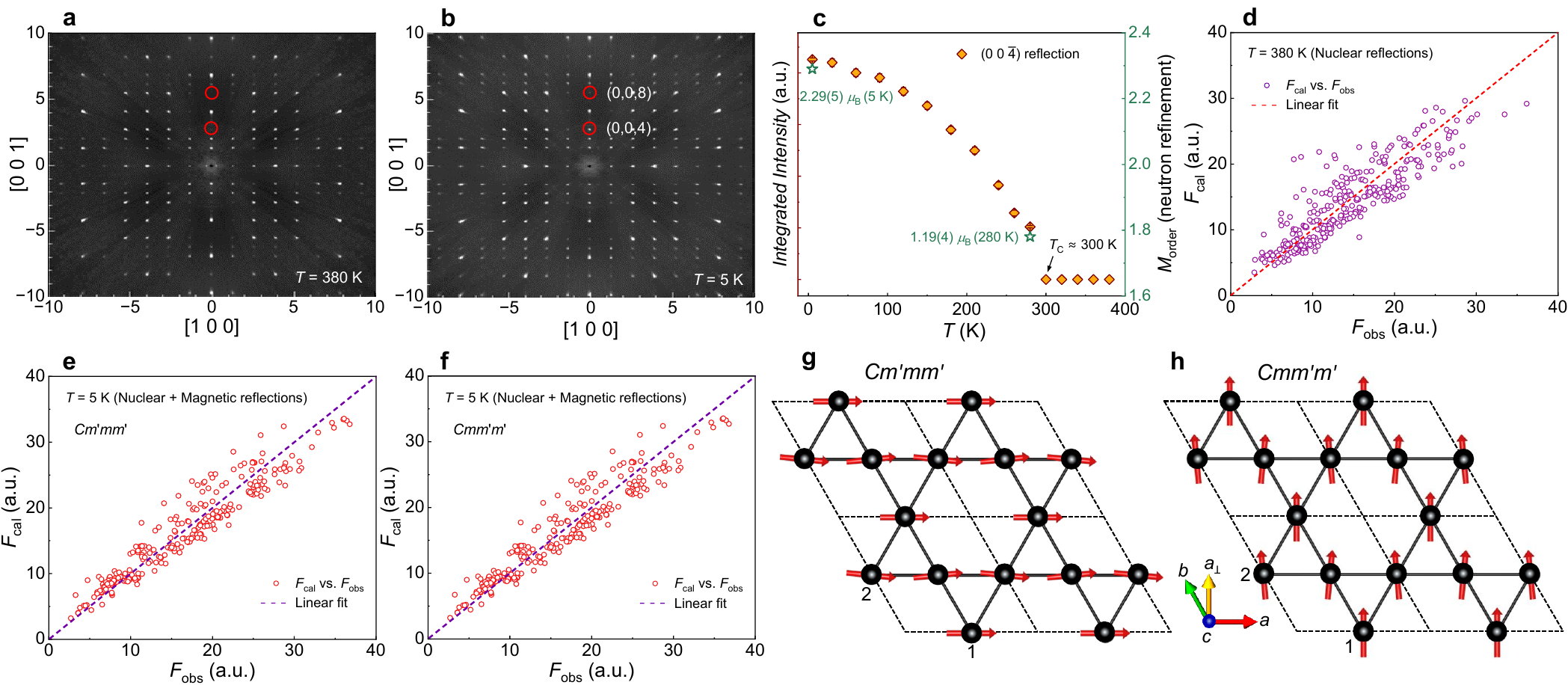}
\caption{
\textbf{a, b}, Integrated intensity maps in the $(h\,0\,l)$ layer of MgMn$_6$Sn$_6$ measured on SXD at $380$~K (\textbf{a}) and 5~K (\textbf{b}). The red circles indicate peaks with a clear change in intensity with temperature. \textbf{c}, Integrated intensity of the magnetic $(0\,0\,\overline{4})$ peak from 380~K down to 5~K, with the Mn ordered moment derived from neutron refinement shown on the right axis. \textbf{e, f}, Plots of calculated vs. experimental magnetic structure factors at 5~K for $Cm'mm'$ (\textbf{e}) and $Cmm'm'$ (\textbf{f}) models. \textbf{g, h}, Corresponding magnetic structures at 5~K for the same models, $Cm'mm'$ (\textbf{g}) and $Cmm'm'$ (\textbf{h}).
}
\label{fig:fig3}
\end{figure*}

To probe the magnetic structure of MgMn$_6$Sn$_6$, we performed SCND experiments over the temperature range 5-380~K. Figs.~3\textbf{a} and~3\textbf{b} present the neutron diffraction patterns measured in the $(h~0~l)$ scattering plane at 380~K and 5~K, respectively. At $T = 380$~K, the diffraction pattern consists exclusively of nuclear reflections allowed by the crystallographic symmetry of the $P6/mmm$ space group. Upon cooling below the magnetic transition temperature (\( T_\mathrm{C} \approx 300~\mathrm{K} \)), a clear enhancement in the intensity of selected Bragg reflections is observed, signaling the onset of long-range magnetic order. These reflections are highlighted by red circles in Figs.~3\textbf{a} and~3\textbf{b}. Notably, no additional purely magnetic reflections appear; instead, all magnetic contributions coincide with the nuclear Bragg peak positions. This behavior indicates a commensurate magnetic structure characterized by a propagation vector $\mathbf{k} = (0,\,0,\,0)$. To further examine the magnetic contribution, we tracked the temperature dependence of selected Bragg reflections, in particular the $(0~0~\overline{4})$ peak (Fig.~3\textbf{c}). The intensity of this reflection increases monotonically upon cooling below $T_C$, consistent with the development of long-range magnetic order. The temperature evolution of the peak intensity closely follows the bulk magnetization $M(T)$ (Fig.~2\textbf{a}), confirming its magnetic origin.

\begin{table*}[t]
\centering
\setlength{\tabcolsep}{22pt}
\setlength{\cmidrulewidth}{0.4pt}

\begin{tabular}{lcccc}

\specialrule{0.4pt}{0pt}{4pt}

Shubnikov space group & \multicolumn{2}{c}{$Cm'mm'$} & \multicolumn{2}{c}{$Cmm'm'$} \\[2pt]

\specialrule{0.4pt}{2pt}{4pt}

& Mn1 & Mn2 & Mn1 & Mn2 \\
\cmidrule(lr){2-5}

Magnetic moment 
& $(m_a,0,0)$ & $(m_a,m_b,0)$ & $(m_a,m_b,0)$ & $(m_a,m_b,0)$ \\

$m_a$ ($\mu_B$)
& $2.29(5)$ & $2.14(6)$ & $1.32(3)$ & $1.10(3)$ \\

$m_b$ ($\mu_B$)
& $0$ & $-0.26(11)$ & $2.65(5)$ & $2.63(5)$ \\

$m_c$ ($\mu_B$)
& $0$ & $0$ & $0$ & $0$ \\

$M$ ($\mu_B$)
& $2.29(5)$ & $2.29(12)$ & $2.29(6)$ & $2.30(3)$ \\

Refinement parameters & & & & \\

GOF
& \multicolumn{2}{c}{2.67} & \multicolumn{2}{c}{2.67} \\

$R_{\mathrm{All}}$
& \multicolumn{2}{c}{9.92} & \multicolumn{2}{c}{9.92} \\

$wR_{\mathrm{All}}$
& \multicolumn{2}{c}{16.76} & \multicolumn{2}{c}{16.76} \\

$R_{\mathrm{Mag}}$
& \multicolumn{2}{c}{7.90} & \multicolumn{2}{c}{7.89} \\

\specialrule{0.4pt}{3pt}{2pt}

\end{tabular}

\caption{Magnetic refinement results from SCND data of MgMn$_6$Sn$_6$ collected at 5~K. $m_a$, $m_b$, and $m_c$ represent the components of the magnetic moment along the crystallographic $a$, $b$, and $c$ axes, respectively, while $M$ denotes the total magnetic moment.}
\label{tab:mag_refinement}

\end{table*}

\par
To determine the symmetry-allowed magnetic structures, a magnetic representation analysis was carried out within the formalism implemented in the \textsc{JANA2020} suite \cite{petvrivcek2023jana2020}. The analysis was performed using the crystallographic information obtained from the Rietveld refinement of the nuclear structure, taking the paramagnetic space group $P6/mmm$ as the parent symmetry. The experimentally established propagation vector $\mathbf{k} = (0, 0, 0)$ was employed, corresponding to magnetic ordering at the Brillouin zone centre. Under this condition, the magnetic representation associated with the magnetic Wyckoff site was decomposed into irreducible representations of the little group of $\mathbf{k}$, yielding the complete set of symmetry-allowed magnetic basis vectors. Each irreducible representation defines a distinct class of magnetic order parameters and generates one or more magnetic space groups (Shubnikov groups), depending on the choice of order parameter direction and the imposed symmetry constraints. This procedure, therefore, enumerates all symmetry-permitted magnetic configurations consistent with the crystallographic symmetry and the propagation vector. The resulting magnetic models, including their corresponding irreducible representations and associated magnetic space groups, are summarized in Table~S2 (SI). Among the $20$ possible magnetic space groups, only two Shubnikov groups, $Cm'mm'$ and $Cmm'm'$, yield physically meaningful solutions with finite magnetic moments. However, refinements based on these two models produce nearly identical reliability factors and comparable Mn moments at both 280~K (Table~S4 and Fig.~S2, SI) and 5~K (Table~II and Figs.~3\textbf{e} and~3\textbf{f}). As a result, the two models cannot be distinguished within the experimental resolution. Both magnetic structures are noncollinear but coplanar, with all Mn moments confined to the basal $(ab)$ plane. The spin arrangement is identical in the two kagome layers within the unit cell. The magnetic structure involves two symmetry-inequivalent Mn sites, denoted Mn1 and Mn2. In the $Cm'mm'$ model, the Mn1 moments align parallel to the crystallographic $a$ axis, while the Mn2 moments are slightly canted away from the $a$ axis within the basal plane (Fig.~3\textbf{g}). In contrast, in the $Cmm'm'$ model, the Mn1 moments are oriented perpendicular to the $a$ axis, and the Mn2 moments are correspondingly canted within the plane (Fig.~3\textbf{h}). The canting of the Mn2 moments is more pronounced at 280~K (Fig.~S2, SI), whereas at 5~K the canting angle is reduced to approximately $5.5^\circ$. Both models are consistent with the bulk $M(T)$ and $M(B)$ measurements, supporting an in-plane ferromagnetic character of MgMn$_6$Sn$_6$.

\par
To further address the ambiguity between these two symmetry-allowed magnetic models, angle-dependent in-plane isothermal magnetization measurements were performed at $T = 5$~K, with the magnetic field applied parallel and perpendicular to the crystallographic $a$ axis (Fig.~S3, SI) on a square crystal to avoid shape anisotropy. A small anisotropy between the two field orientations is observed; however, the difference in the saturation behavior is minimal and lies within the experimental uncertainty. Therefore, these measurements do not provide a definitive distinction between the $Cm'mm'$ and $Cmm'm'$ magnetic models. Both configurations remain consistent with the neutron diffraction data as well as the bulk magnetization results.
\par
In MgMn$_6$Sn$_6$, the Mn moments are coupled via itinerant conduction electrons, leading to exchange interactions that can be described within an RKKY-type framework~\cite{biswas2025tuning}. Due to the oscillatory and long-range nature of the RKKY interaction, competing intraplanar and interplanar Mn-Mn exchange interactions are expected. Previous first-principles calculations, limited to the collinear magnetic configurations, have reported nearly degenerate ferromagnetic (FM) and antiferromagnetic (A-type) states~\cite{chen2025computational}. This close energetic proximity suggests that the magnetic ground state is highly sensitive to subtle perturbations and competing interactions~\cite{sau2024exploring}. Under these conditions, the system can lower its energy through a small canting of spins, resulting in a noncollinear magnetic arrangement. The weak noncollinearity observed in the present SCND measurements is consistent with this scenario.

\begin{figure*}[t]
\centering
\includegraphics[width=1\textwidth]{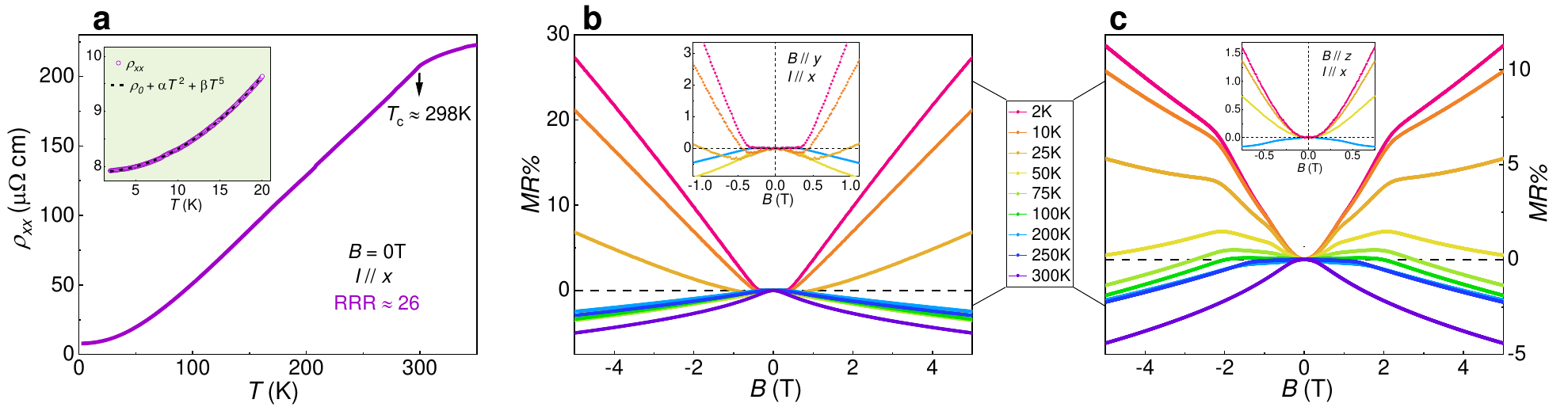}
\caption{\textbf{a}, Electrical resistivity as a function of temperature for current applied along the x direction and the inset shows the fitting of the resistivity curve at low temperatures. \textbf{b}, \textbf{c}, Magnetoresistance vs. B at selected temperatures, measured with the magnetic field along the y (\textbf{b}) and z (\textbf{c}) directions, respectively. The insets show enlarged views of the low-field region for both field orientations.}
\label{fig:fig4}
\end{figure*}

\subsection{Resistivity and magnetoresistance}%
The temperature-dependent in-plane resistivity $\rho(T)$ at 0~T, measured with $I \parallel x$, is shown in Fig.~4\textbf{a}. At 2~K, the residual resistivity ($\rho_{0}$) is 7.9~$\mu\Omega$-cm, giving a high residual resistivity ratio ($RRR$ = $\rho_{xx}$(300~K)/$\rho_{xx}$(2~K)$ \approx 26$), which reflects the good quality of as-grown single crystals. A clear anomaly in $\rho(T)$ appears at 298~K, indicative of the onset of a FM transition, as corroborated by magnetic measurements. Below this temperature, the resistivity decreases smoothly, following a metallic trend down to 2~K. To elucidate the dominant scattering mechanisms, the low-temperature resistivity data (up to 20~K) were fitted using the expression $\rho(T) = \rho_0 + aT^2 + bT^5$ (Inset of Fig.~4\textbf{a}), yielding fitting coefficients $a = 0.42 \times 10^{-2}~\mu\Omega\,\mathrm{cm\,K^{-2}}$ and $b = 0.19 \times 10^{-7}~\mu\Omega\,\mathrm{cm\,K^{-5}}$. The $T^2$ term is characteristic of electron-electron (Fermi-liquid) scattering, whereas the $T^5$ contribution arises from electron-phonon scattering in the low-temperature regime. It is evident that, at low temperatures, the resistivity is predominantly governed by the electron-electron scattering mechanism.

To investigate the magnetoresistance (MR) effect, the magnetic-field dependence of the longitudinal resistivity was measured at various temperatures. The MR is defined as $[\rho(B) - \rho(0)]/\rho(0) \times 100\%$. Figs.~4\textbf{b} and~4\textbf{c} show the MR for magnetic fields applied along the $y$- and $z$-directions, respectively, with the current flowing along the $x$-direction. For $B \parallel y$, the MR of MgMn$_6$Sn$_6$ exhibits two distinct temperature-dependent regimes. At low temperatures ($T < 50$~K), the MR is initially negative at low magnetic fields and reaches a minimum near the saturation field $B_s$. This behavior arises from the field-induced suppression of spin-disorder scattering in the regime where magnetic domains progressively align with the applied field. Upon increasing the magnetic field beyond $B_s$, the MR curve turns upward and increases approximately linearly with field, indicating that the conventional orbital magnetoresistance dominates once the magnetization is saturated. At 2~K, the MR reaches about 27\% at an applied field of 5~T. In contrast, at elevated temperatures, the MR remains negative over the entire measured field range. In this regime, enhanced thermal spin fluctuations strengthen the spin-disorder contribution, while the positive orbital MR is progressively reduced due to decreased carrier mobility, resulting in a net negative MR.

For $B \parallel z$, corresponding to the magnetic hard axis, the MR remains positive at low temperatures even in the low-field regime. Below the saturation field $B_s \approx 2$~T, the MR exhibits an approximately quadratic field dependence, consistent with dominant orbital magnetoresistance. In contrast to the in-plane field configuration, a hard-axis field leads to only a gradual reduction of magnetic scattering below $B_s$ for in-plane transport, allowing the orbital contribution to dominate and yield a positive MR. Above $B_s$, the MR continues to increase smoothly with field, consistent with an orbital-dominated transport regime, and at higher temperatures it follows a similar overall trend to that observed for $B \parallel y$.

\subsection{Hall measurements}%
To investigate the electronic properties of MgMn$_6$Sn$_6$, Hall measurements were performed over a broad temperature range from 2~K up to 300~K using two different configurations: $\rho_{zx}$ ($I \parallel x$ and $B \parallel y$) and $\rho_{yx}$ ($I \parallel x$ and $B \parallel z$), as shown in the Figs.~5\textbf{a} and~5\textbf{b}. The Hall resistivity for both field orientations exhibits a pronounced dependence on magnetization, a hallmark of the anomalous Hall effect (AHE) in FM conductors. In this context, the Hall resistivity can be described as $\rho_{ji} = R_0 B + R_S \mu_0 M = \rho_{ji}^N + \rho_{ji}^A$, where $R_0$ is the normal Hall coefficient (arising from the Lorentz force) and $R_S$ is the anomalous Hall coefficient. Here, $i, j \in \{x, y, z\}$, $\rho_{ji}^N$ represents the normal Hall resistivity, and $\rho_{ji}^A$ denotes the anomalous Hall resistivity (AHR) \cite{nagaosa2010anomalous}. Figs.~5\textbf{a} and 5\textbf{b} show that the Hall resistivity is relatively weak at low temperatures and increases progressively upon warming. The AHR, $\rho_{ji}^{\mathrm{A}}$, is extracted by linearly fitting the high-field regime of the Hall resistivity to isolate the normal Hall component, and extrapolating the fit to zero magnetic field; the intercept at zero field yields $\rho_{ji}^{\mathrm{A}}$. The Hall conductivity is calculated using the relation $\sigma_{ij} = \rho_{ji} / (\rho_{ji}^2 + \rho_{ii}^2)$. A similar procedure is applied to the field-dependent $\sigma_{ij}$ data to extract the AHC. The AHC, $\sigma_{ij}^{\mathrm{A}}$, originates from both intrinsic and extrinsic mechanisms, i.e., the Berry curvature-driven Karplus-Luttinger mechanism and extrinsic contributions namely skew scattering and side jump \cite{karplus1954hall,nagaosa2010anomalous,smit1955spontaneous,smit1958spontaneous,berger1970side}. Among these, the intrinsic contribution is purely determined by the electronic band structure and is expected to be largely temperature-independent \cite{nagaosa2010anomalous}.

Now, for the in-plane magnetic field ($B \parallel y$) configuration the AHR ($\rho_{zx}^{\mathrm{A}}$) increases monotonically with increasing temperature, while the corresponding Hall conductivity ($\sigma_{xz}^{\mathrm{A}}$) reaches a maximum value of $2200 \pm 65.47~(\Omega\cdot\mathrm{cm})^{-1}$ at 2~K, followed by a sharp decline with temperature, becoming nearly temperature-independent above 50~K (Fig.~5\textbf{c}). In contrast, for out-of-plane magnetic field ($B \parallel z$) configuration (Fig.~5\textbf{d}), the AHE exhibits a markedly different behavior. The inset of Fig.~5\textbf{d} shows that $\rho_{yx}^{\mathrm{A}}$ undergoes a clear sign reversal just above 10 K, a behavior not observed for the in-plane field configuration. Consistently, the AHC $\sigma_{xy}^{\mathrm{A}}$ changes abruptly from a negative value of $-268.76 \pm 13.71~(\Omega\cdot\mathrm{cm})^{-1}$ at 2 K to a large positive value of $380.94 \pm 8.10~(\Omega\cdot\mathrm{cm})^{-1}$ at 25 K. Above $\sim$25 K, $\sigma_{xy}^{\mathrm{A}}$ remains positive and decreases gradually with further temperature increase (Fig.~5\textbf{d}).

\begin{figure*}[t]
    \centering
    \includegraphics[width=1\textwidth]{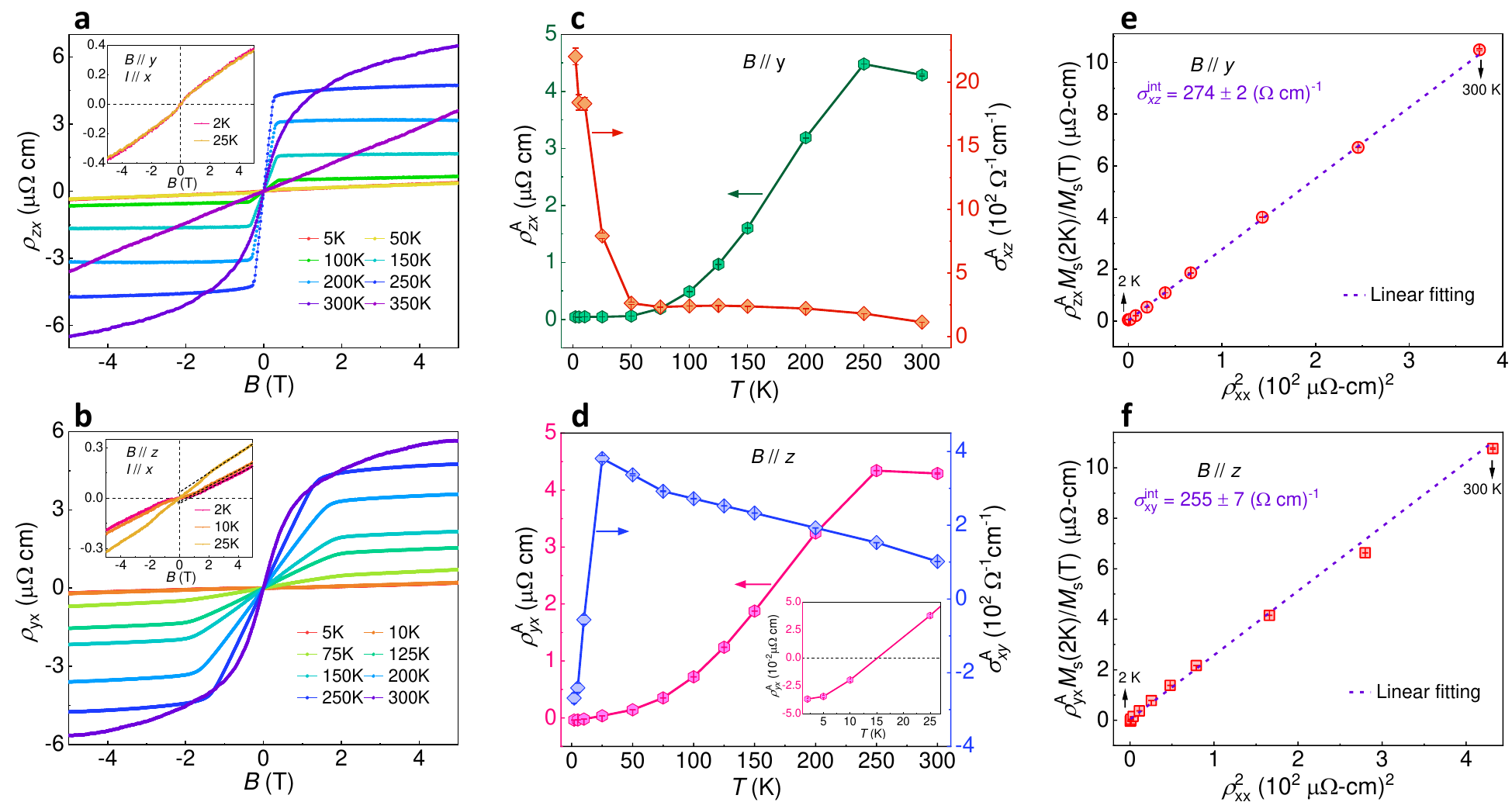}
    \caption{\textbf{a}, \textbf{b}, Hall resistivity $\rho_{zx}$ (\textbf{a}) and $\rho_{yx}$ (\textbf{b}) as a function of magnetic field $B$ at selected temperatures. The insets show the corresponding resistivity plots at low temperatures for clarity, and the dashed lines in the inset of \textbf{b} represent the linear fit to $\rho_{yx}^{\mathrm{N}}$. \textbf{c}, \textbf{d}, Temperature dependence of the anomalous Hall resistivity and anomalous Hall conductivity for $B \parallel y$ (\textbf{c}) and $B \parallel z$ (\textbf{d}). \textbf{e}, \textbf{f}, $\rho_{ji}^{\mathrm{A}} M_s(2\,\mathrm{K})/M_s(T)$ plotted against $\rho_{ii}^2$ from 2 to 300~K. $\sigma_{ij}^{\mathrm{int}}$ is obtained from the linear fitting (red dashed line) of the data points, where $i = x$ and $j = y$ (\textbf{e}) and $z$ (\textbf{f}).
  }
    \label{fig:fig5}
\end{figure*}

This pronounced anisotropy in the temperature evolution of the AHE for different applied field directions highlights the directional sensitivity of the Hall response. To elucidate whether the AHC is primarily governed by intrinsic or extrinsic mechanisms, we performed a detailed scaling analysis, discussed in the following section. 
Since the saturation magnetization (\( M_{\text{s}} \)) changes significantly over the temperature range we examine (i.e., \( M_{\text{s}}(300\,\text{K})/M_{\text{s}}(2\,\text{K}) \approx 0.5 \)), we have taken this variation into account in our analysis of the anomalous Hall effect (AHE). To extract the intrinsic contribution to the AHC, we use the following analytical expression \cite{tian2009proper}:
\begin{equation}
\rho_{ji}^{A}  M_{\text{s}}(2\,\text{K}) = \left( \sigma^{\text{int}} \rho_{ii}^2 + \beta^{\text{skew}} \rho_{ii} \right) M_{\text{s}}(T)
\end{equation}

where $\beta^{\text{skew}}$ denotes the skew scattering coefficient, and $\sigma^{\text{int}}$ represents the estimated intrinsic AHC, which may also include a contribution from the side-jump mechanism. By performing a linear fit (shown in Figs.~5\textbf{e} and 5\textbf{f}) of the experimental data, where $\rho_{ji}^{A} M_{\text{s}}(2,\text{K}) / M_{\text{s}}(T)$ is plotted against $\rho_{ii}^2$, we extract the intrinsic conductivities as $\sigma_{xy}^{\text{int}} = (255\pm7)~(\Omega\cdot\text{cm})^{-1}$ and $\sigma_{xz}^{\text{int}} = (274\pm2)~(\Omega\cdot\text{cm})^{-1}$ which are independent of the scattering rate, consistent with a system possessing significant Berry curvature contributions. Therefore, the intrinsic AHC, $\sigma_{xy}^{\mathrm{int}}$, of MgMn$_6$Sn$_6$ is estimated to be $\sim 0.29~e^2/h$ per kagome layer. This value is comparable to those reported for other members of the rare-earth based $R$Mn$_6$Sn$_6$ family, such as $\sim 0.33~e^2/h$ for Sm \cite{PhysRevB.103.235109} and $\sim 0.27~e^2/h$ for Gd \cite{asaba2020anomalous}, while being modestly smaller than that of LiMn$_6$Sn$_6$ ($\sim 0.44~e^2/h$) \cite{PhysRevB.103.144410}.

The nearly isotropic intrinsic AHC suggests that the Berry curvature distribution, and hence the electronic structure near $E_F$, is relatively insensitive to magnetic field direction. Interestingly, the temperature-dependent AHC exhibits a clear upturn in $\sigma_{xz}^{\mathrm{A}}$ ($B \parallel y$) and a pronounced downturn in $\sigma_{xy}^{\mathrm{A}}$ ($B \parallel z$) at low temperatures (Figs.~5\textbf{e} and 5\textbf{f}). This contrasting behavior may be indicative of the interplay between intrinsic and extrinsic contributions to the anomalous Hall effect. While the nearly isotropic values of intrinsic AHC extracted at intermediate temperatures point to dominant Berry curvature effects, the low-temperature evolution likely reflects the increasing influence of extrinsic mechanisms, i.e., skew scattering and/or side-jump. The side jump is generally proportional to $(e^2/hc)(E_{\mathrm{soc}}/E_F)$, where $e$ = electronic charge, $h$ = Plank constant, $c$ =cross-plane lattice constant and  $E_{\mathrm{soc}}$ = spin-orbit coupling energy. In typical FM metals, $E_{\mathrm{soc}}/E_F$ ranges from 0.1 to 0.001, indicating that the side-jump contributions are typically small compared to the total AHC \cite{berger1970side,nagaosa2010anomalous}. The skew-scattering contribution can be expressed as $\sigma_{ij}^{\mathrm{skew}} = \frac{2e^{2}}{hc}\,\frac{E_{F}\tau}{\hbar}\,S$, where $\tau$ is the carrier lifetime and $S \sim E_{\mathrm{soc}} v_{\mathrm{imp}}/W^{2} \,(\ll 1)$ is the skewness factor determined by the impurity potential $v_{\mathrm{imp}}$ and the electronic bandwidth $W$ (inverse of the density of states) \cite{onoda2006intrinsic}. At low temperatures the system lies predominantly in the clean-limit regime, where the suppression of phonon scattering increases the carrier lifetime $\tau$, thereby enhancing the skew-scattering contribution to the AHC. Therefore, in \(\mathrm{MgMn_6Sn_6}\), the contrasting low temperature behavior is most likely driven by skew scattering effect. In this regime the observed sign reversal suggests that skew scattering may contribute with opposite polarity along different crystallographic directions, highlighting the anisotropic nature of extrinsic contribution of AHE in this kagome system.

\begin{figure}[t]
\centering
\includegraphics[width=\linewidth]{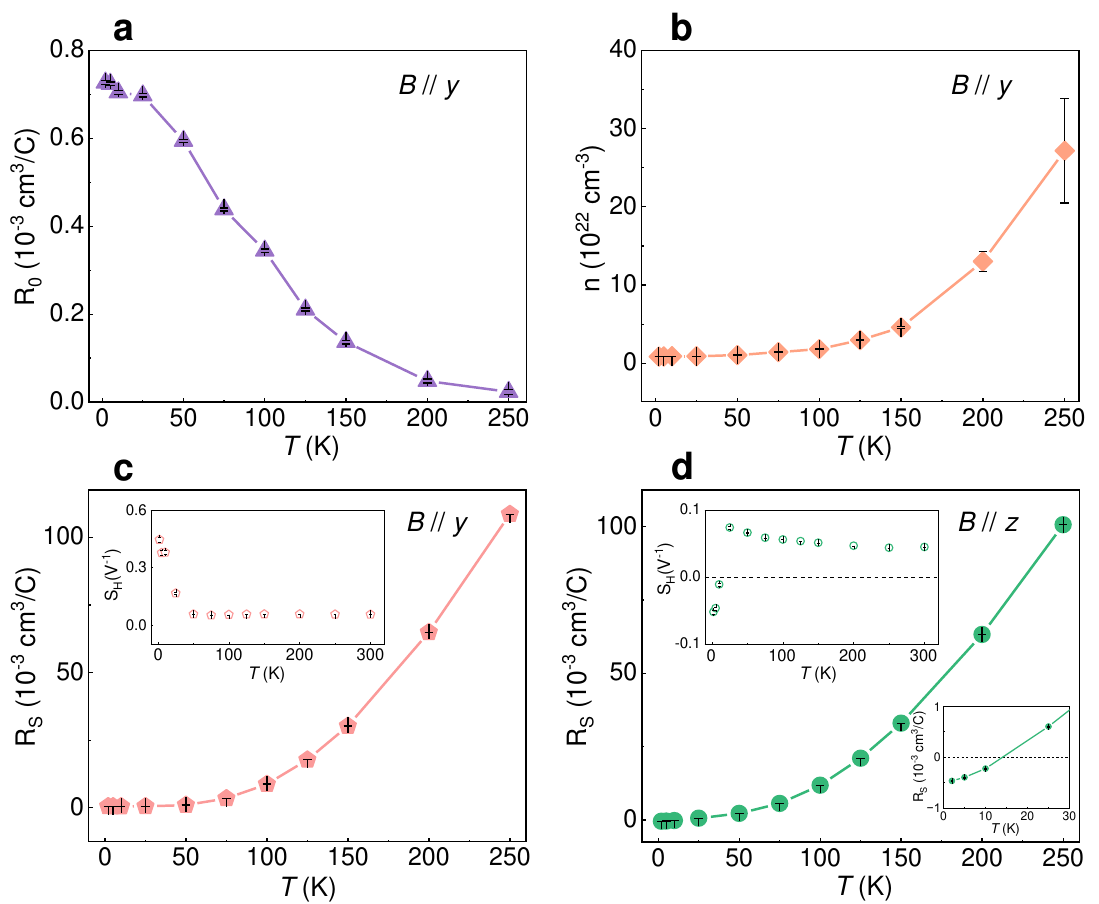}
\caption{\textbf{a}, \textbf{b}, Temperature dependence of the normal Hall coefficient $R_0$ (\textbf{a}) and the corresponding carrier density $n$ (\textbf{b}) for $B \parallel y$. \textbf{c}, \textbf{d}, Anomalous Hall coefficient $R_s$ plotted as a function of temperature for $B \parallel y$ (\textbf{c}) and $B \parallel z$ (\textbf{d}). The top-left insets in \textbf{c} and \textbf{d} show the corresponding temperature dependence of the anomalous Hall scaling factor. The bottom-right inset in \textbf{d} highlights $R_s$ in the low-temperature regime.}
\label{fig:fig6}
\vspace{-0.4cm}
\end{figure}

Additionally, $R_0$ and $R_s$ were extracted from linear fits to $\rho_{ji}/B = R_0 + R_s \mu_0 M/B$ for both magnetic-field orientations. The positive sign of $R_0$ over the entire temperature range indicates that charge transport in $\mathrm{MgMn_6Sn_6}$ is hole-dominated (Figs.~6\textbf{a} and S4\textbf{c}). Within a single-band approximation, the carrier density is estimated as $n = 1/(|e|R_0)$, yielding $n \approx 0.89 \times 10^{22}~\mathrm{cm}^{-3}$ at 10~K for $B \parallel y$, corresponding to approximately 2.1 carriers per formula unit. For both field orientations, $R_0$ decreases while $R_s$ increases with increasing temperature, indicating that the Hall response at low temperatures is dominated by the normal Hall contribution. Notably, for $B \parallel z$, $R_s$ undergoes a sign reversal above 10~K (lower inset of Fig.~6\textbf{d}), which can be attributed to the crossover behavior of $\rho_{yx}^{\mathrm{A}}$. To further examine the scaling behavior of the AHC, we evaluate $S_H = |\sigma_{ij}^{\mathrm{A}}|/M$, which characterizes the coupling strength between the anomalous Hall response and magnetization. For both field orientations, $S_H$ remains nearly temperature-independent above 25~K (insets of Figs.~6\textbf{c} and 6\textbf{d}), consistent with an intrinsic mechanism of the anomalous Hall effect. The pronounced variation observed below $\sim 25$~K coincides with changes in the magnitude of $|\sigma_{ij}^{\mathrm{int}}|$ (see Figs.~5\textbf{c} and 5\textbf{d}). The extracted $S_H$ values are comparable to those reported for conventional itinerant ferromagnets such as Fe and Ni, typically in the range $0.01$-$0.2~\mathrm{V}^{-1}$~\cite{nagaosa2010anomalous,PhysRevB.94.075135}.

\par
Beyond the electronic transport analysis discussed above, we do not find any evidence of THE in this compound. Although the magnetic ground state is non-collinear, the spin texture does not develop a finite non-coplanar component under applied magnetic field, and therefore no measurable SSC emerges. As a result, the Hall response in MgMn$_6$Sn$_6$ is fully governed by the normal and anomalous contributions. The microscopic origin of the anomalous Hall effect was investigated using first-principles calculations.

\begin{figure}[t]
\centering
\includegraphics[width=\linewidth]{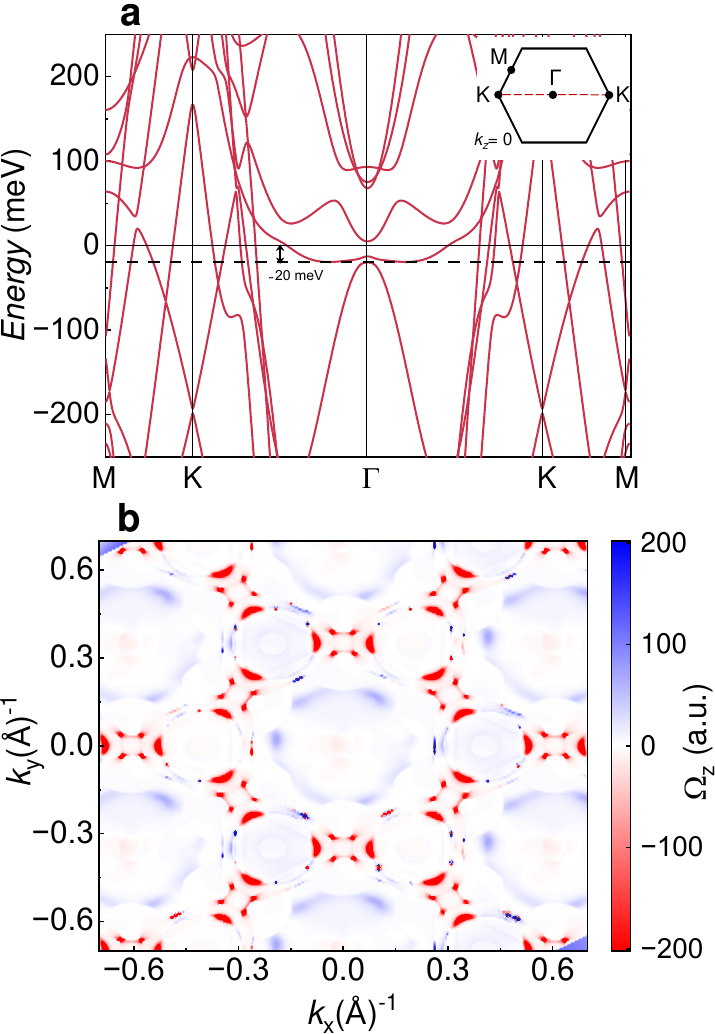}
\caption{\textbf{a}, Calculated electronic band structure of MgMn$_6$Sn$_6$ with SOC along the high-symmetry path $M$–$K$–$\Gamma$–$K$–$M$. The horizontal black dashed line indicates a downward shift of $E_F$ by $\sim 20$~meV used to reproduce the experimental AHC value. Inset: First BZ of $\mathrm{MgMn_6Sn_6}$ with the high-symmetry points marked. \textbf{b}, Berry curvature distribution in the $k_z = 0$ plane.}
\label{fig:fig7}
\vspace{-0.4cm}
\end{figure}
  
\subsection{First-principles calculations}
We now study the electronic structure of this material using the ab-initio DFT calculations, assuming a collinear spin configuration within the PBE$+U$+SOC framework to understand the magnetic and transport properties. The spin-resolved total and orbital-projected densities of states (DOS), shown in SI, Fig.~S5\textbf{a}, indicate that Mn is the dominant contributor to the magnetic moment, whereas Sn states contribute negligibly to the net magnetization. The calculations yield a magnetic moment of 2.3~$\mu_{\mathrm{B}}$/Mn, which is in fair agreement with the experimentally observed value.

Owing to the presence of kagome and honeycomb sublattices, $\mathrm{MgMn_6Sn_6}$, naturally hosts a variety of symmetry-protected electronic features. The calculated band dispersions in the $k_z = 0$ plane along the high-symmetry path $M$-$K$-$\Gamma$-$K$-$M$ are shown in Fig.~7\textbf{a} (with SOC) and in SI, Fig.~S5\textbf{b} (without SOC). A set of linear band crossings appears slightly below $E_F$, including Weyl points located approximately 44~meV and 67~meV below $E_F$ along the $\Gamma$-$K$-$M$ line. At the $K$ point, we also identify a linear crossing point located $\sim$200~meV below $E_F$ due to the combined effect of the two-fold ($C_{2x}$) and three-fold ($C_{3z}$) rotational symmetries~\cite{ye2018massive} of the kagome network, which develops a small gap of 5.5~meV on the application of SOC. Orbital-projected analysis reveals that this crossing point is predominantly derived from the out-of-plane Mn $d_{z^2}$ orbital. Another notable feature is the touching of a nearly flat band with a quadratic dispersive band at the $\Gamma$-point in the absence of SOC (see SI, Fig.~S5\textbf{b}). The inclusion of SOC opens a gap of 7 meV at the $\Gamma$ point. The features in the band structure discussed above have the potential to generate large Berry curvature (BC). 

In Fig.~7\textbf{b}, we present the  BC distribution in the $k_z = 0$ plane of the Brillouin zone (BZ), evaluated at $E_F$. The BC landscape exhibits distinct regions of opposite sign, with red and blue shading denoting its extrema. A clear three-fold rotational symmetry ($C_{3z}$), dictated by the kagome lattice geometry, is prominently preserved. Strikingly, pronounced negative BC ``hot spots'' emerge near specific high-symmetry points, suggesting their dominant role in governing transverse transport. As reported by us in Ref.~\cite{sau2024exploring} the energy dependence of the AHC reaches a value of approximately 500~$(\Omega\cdot\mathrm{cm})^{-1}$ at $E_F$, which is significantly larger than the experimentally observed value of $\sim 255~(\Omega\cdot\mathrm{cm})^{-1}$. The experimental magnitude can be reproduced by shifting $E_F$ downward by about 20~meV, which may originate from slight off-stoichiometry due to self-doping during single crystal growth. Notably, this downward shift moves $E_F$ deeper into the hole-like bands near the $\Gamma$ point (see black dashed line in Fig.~7\textbf{a}), which is consistent with the experimentally observed hole-dominated electrical transport in MgMn$_6$Sn$_6$.

\subsection{Heat capacity}

\begin{figure}[t]
\centering
\includegraphics[width=\linewidth]{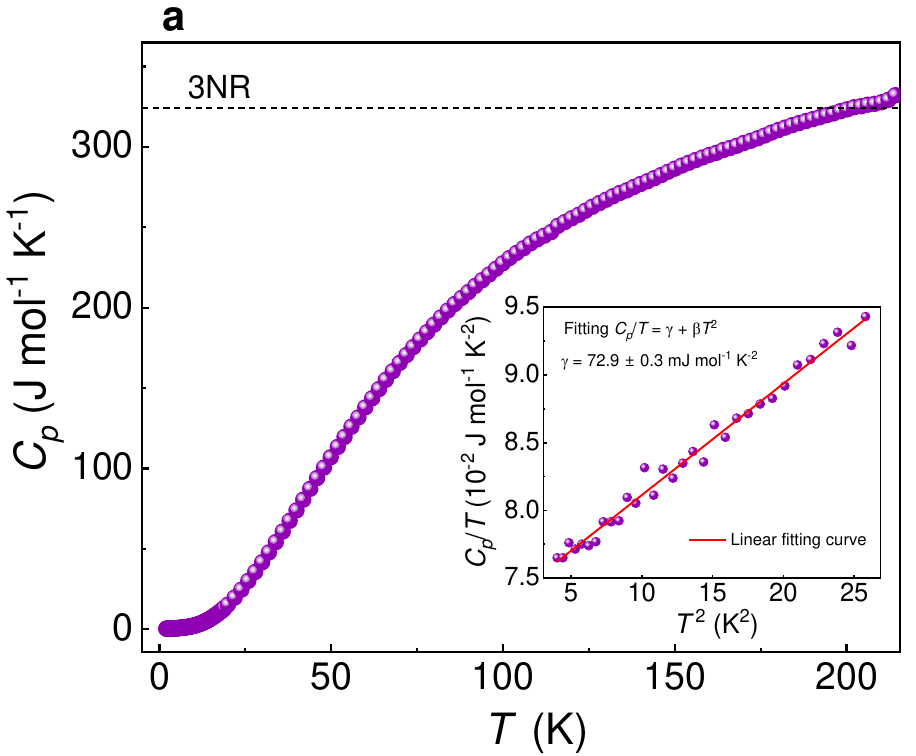}
\caption{\textbf{a}, Specific heat $C_p(T)$ as a function of temperature for a MgMn$_6$Sn$_6$ single crystal. Inset: $C_p/T$ versus $T^2$; the solid red line represents the linear fit to the data in the temperature range 2-5~K.}
\label{fig:fig8}
\vspace{0.3cm}
\end{figure}

To probe the thermodynamic properties of MgMn$_6$Sn$_6$, heat capacity measurements were carried out, and the results are shown in Fig.~8\textbf{a}. At high temperatures, the lattice contribution to the heat capacity of a solid approaches the classical Dulong-Petit limit, $3NR = 324.25$ J mol$^{-1}$ K$^{-1}$, where $N = 13$ is the number of atoms per formula unit and $R$ is the universal gas constant. However, above $\sim$200 K  the additional contributions beyond the lattice term are mainly associated with the magnetic effect as the temperature approches to the transition temperature. The low-temperature data were fitted using the Debye model, $C = \gamma T + \beta T^3$, where the linear term represents the electronic contribution and the cubic term corresponds to lattice (phonon) vibrations (see inset of Fig.~8\textbf{a}). The linear dependence of $C_p/T$ on $T^2$ confirms the validity of this model. The fitting yields a Sommerfeld coefficient $\gamma = 72.9 \pm 0.3$ $\mathrm{mJ\,mol^{-1}\,K^{-2}}$ and $\beta = 0.82 \pm 0.02$ $\mathrm{mJ\,mol^{-1}\,K^{-4}}$, with a goodness of fit $R^2 = 0.9864$. The Debye temperature $\Theta_D$ is calculated using the relation $\Theta_D = \left( {12 \pi^4 N R}/{5 \beta} \right)^{1/3}$, giving $\Theta_D = 313.0 \pm 2.2$ K.

\begin{table}[t]
\centering
\begin{tabular}{lcc}
\specialrule{0.4pt}{0pt}{4pt}
Compound & $\gamma$ (mJ\,mol$^{-1}$\,K$^{-2}$) & Ref. \\
\specialrule{0.4pt}{2pt}{4pt}
LuFe$_6$Sn$_6$  & 87 & \cite{lyu2024anomalous}  \\
YFe$_6$Ge$_6$   & 84 & \cite{nayak2025crystalgrowthphysicalproperties} \\
YV$_6$Sn$_6$    & 67 & \cite{pokharel2021electronic}  \\
YCr$_6$Ge$_6$   & 66 & \cite{ishii2013ycr6ge6}  \\
LuCr$_6$Ge$_6$  & 57.3 & \cite{riedel2025structural}  \\
LuMn$_6$Sn$_6$  & 48 & \cite{li2024enhanced}  \\
YMn$_6$Sn$_6$   & 42.9 & \cite{wang2023magnetic}  \\
ThV$_6$Sn$_6$   & 41.7 & \cite{xiao2024preparation}  \\
MgCo$_6$Ge$_6$  & 20.8 & \cite{sinha2021twisting}  \\
LuV$_6$Sn$_6$   & 17.4 & \cite{lv2026cooperative}  \\
\specialrule{0.4pt}{2pt}{4pt}
YbV$_6$Sn$_6$   & 411 & \cite{guo2023triangular} \\
YbCr$_6$Ge$_6$  & 203 & \cite{lv2026cooperative} \\
YbMn$_6$Sn$_6$  & 127 & \cite{li2024enhanced} \\
YbFe$_6$Ge$_6$  & 90 & \cite{avila2005direct} \\
UCr$_6$Ge$_6$   & 86.5 & \cite{riedel2025structural} \\
GdNb$_6$Sn$_6$  & 85.1 & \cite{xiao2025kagome} \\
TbV$_6$Sn$_6$   & 72.1 & \cite{pokharel2022highly} \\
UV$_6$Sn$_6$    & 40 & \cite{thomas2025unusual} \\
\specialrule{0.4pt}{3pt}{2pt}
\end{tabular}
\caption{Sommerfeld coefficient $\gamma$ for reported $RT_6X_6$ compounds. Compounds listed above the mid horizontal line correspond to $d$-electron systems, whereas those below include additional $f$-electron contributions.}
\label{tab:gamma_comparison}
\end{table}

Notably, the obtained $\gamma$ is relatively large compared to many other $d$-electron and $d$+$f$-electron based $R166$ compounds, as summarized in Table~IV. The DOS at $E_F$, $D(E_F)$, estimated from $\gamma$ using $D(E_F) = {3\gamma}/{\pi^2 k_B^2 N_A}$, where $k_B$ is the Boltzmann constant and $N_A$ is Avogadro’s number, is found to be 30.9~$\mathrm{states\,eV^{-1}\,f.u.^{-1}}$, while the DFT DOS value is 12.7~$\mathrm{states\,eV^{-1}\,f.u.^{-1}}$ (see Fig.~S5\textbf{a}; SI). This more than twofold enhancement in $\gamma$ suggests an increased quasiparticle effective mass. Notably, there has been reports of heavy-fermion-like behavior in non \textit{f}-electrons systems which mainly stems from mechanisms such as localized-itinerant crossover or geometrical frustration \cite{cheng2013possible,laad2003heavy,kotegawa2020helimagnetic}. In MgMn$_6$Sn$_6$ however, the large enhancement of $\gamma$ most likely points toward the presence of nearly flat bands in the vicinity of $E_F$, as observed in our band structure calculations. Detailed angle-resolved photoemission or scanning tunneling spectroscopy can verify this experimentally.

\section{\label{sec:level3}CONCLUSION} 

In summary, by combining SCND, magneto-transport, and heat capacity measurements, we present a comprehensive investigation of the ground-state magnetic structure and electronic transport properties of the room-temperature kagome ferromagnetic metal MgMn$_6$Sn$_6$. SCND results reveal that the Mn moments form a coplanar yet non-collinear spin arrangement within the kagome bilayer, characterized by a canting angle of $\sim 5.5^{\circ}$ at 5~K, which gradually increases with increasing temperature. This non-collinear spin configuration suggests the presence of competing exchange interactions between the nearest- and next-nearest-neighbor Mn atoms within and between the kagome layers, which calls for further theoretical analysis to quantitatively determine the underlying exchange parameters. Hall effect measurements reveal an intrinsic AHC of $\sim 0.29\,e^{2}/h$ per kagome layer, which remains nearly independent of the applied magnetic field direction, consistent with a Berry-curvature-driven mechanism as supported by our DFT calculations. At low temperatures, however, a crossover from intrinsic to extrinsic contributions is observed. In this regime, the AHC exhibits opposite trends depending on the field orientation, indicating a significant skew-scattering contribution with opposite polarity for different field directions. This behavior points toward anisotropic skew scattering associated with the underlying electronic structure and scattering processes in MgMn$_6$Sn$_6$. A large value of Sommerfeld coefficient ($\gamma$ $\approx$ 73 $\mathrm{mJ\,mol^{-1}\,K^{-2}}$) suggests strong electron correlation effect in this system which further supports the presence of kagome-driven flat band near $E_F$. Our findings suggest that the room temperature non-collinear ferromagnetic kagome MgMn6Sn6 exhibits characteristic features of a correlated topological metal, which can be further tuned through chemical substitution and pressure.

\section{\label{sec:level4}METHODS} 
\textit{Experimental details.} 
Single crystals of MgMn$_6$Sn$_6$ were synthesized using the self-flux method \cite{song2024magnetocaloric}. Magnesium turnings (99.9\%), manganese pieces (99.9\%), and tin pieces (99.98\%) were used as starting materials in an atomic ratio of Mg:Mn:Sn = 4:1:8. The elements were loaded into an alumina crucible inside a nitrogen-filled glove box (oxygen and moisture levels $<$0.1~ppm), and the crucible was sealed in an evacuated quartz tube. The sealed quartz tube was heated to 500 °C at a rate of 30 °C/h, held at this temperature for 5 hours, and then heated to 800°C at the same rate. It was maintained at 800 °C for 10 hours before being cooled to 420 °C at a rate of 4 °C/h. At 420 °C, the residual Sn flux was separated from the crystals by centrifugation. The obtained crystals exhibit a hexagonal, plate-like morphology with lateral dimensions of 1-3 mm and thickness ranging from 0.1 to 0.7 mm. The single-crystalline nature was confirmed through X-ray diffraction (XRD) measurements on different crystal facets, as shown in Fig.~1\textbf{d}. The insets of this figure display the crystals used for the respective XRD measurements. The Laue backscattering pattern (left inset of Fig.~1\textbf{d}), recorded with the beam along the $c$-axis, shows sixfold symmetry, confirming high crystalline quality. The chemical compositions of MgMn$_6$Sn$_6$ were verified by energy-dispersive X-ray spectroscopy (EDXS, Quanta 250 FEG SEM). Representative EDXS compositions and elemental maps are provided in SI (Fig.~S1 and Table~S1). Room-temperature single-crystal X-ray diffraction (SCXRD) and powder X-ray diffraction (PXRD) measurements were performed to determine the crystal structure and confirm phase purity. PXRD was carried out on crushed crystals of MgMn$_6$Sn$_6$ using the Rigaku SmartLab diffractometer with 9 kW Cu K$_\alpha$ ($\lambda$ = 1.5418 Å) radiation. FULLPROF software was utilized to refine the XRD reflection patterns.

\par
SCXRD data were collected from a sample with dimensions of $0.05 \times 0.067 \times 0.195~\mathrm{mm}^3$ using a Rigaku Oxford Diffraction XtaLAB Synergy diffractometer equipped with a HyPix 6000 photon-counting detector and Mo radiation. A full sphere of data was collected as a series of $\omega$ scans at fixed $\phi$ and $\kappa$ settings, with an exposure time of 0.14~s per frame. The data were processed using \textsc{CrysAlisPro}, including absorption correction based on face-indexing and numerical Gauss integration \cite{agilent2014agilent}.

\par
SCND measurements were performed on the Single Crystal Diffractometer (SXD) beamline at the ISIS spallation neutron source of Rutherford Appleton Laboratory, UK. The SXD instrument utilizes the time-of-flight Laue technique, which enables access to a large three-dimensional (3D) volume of reciprocal space within a single measurement \cite{SXD}. For the experiment, a single crystal of approximate dimensions $2 \times 3 \times 0.2$~mm$^{3}$ (mass $\approx 3$~mg) was mounted on an aluminium (Al) pin using thin strips of Al adhesive tape. The sample was placed inside an evacuated sample environment and cooled using a closed-cycle refrigerator. Diffraction data were collected over the temperature range from 5~K to 380~K for several fixed crystal orientations. Data reduction was carried out using the SXD2001 software package \cite{SXD2001}. Only reflections satisfying $I > 3\sigma(I)$ were included in the refinements, where $I$ and $\sigma(I)$ denote the measured intensity and its standard deviation. Both the nuclear and magnetic structure refinements were carried out using the \textsc{JANA2020} software package \cite{petvrivcek2023jana2020}, and an isotropic Becker-Coppens extinction (type-II) was employed to refine the extinction parameters \cite{becker1974extinction}.
\par
The magnetic measurements were conducted on polished crystals using the Vibrating Sample Magnetometer (VSM) option of the Quantum Design Physical Property Measurement System (DynaCool, PPMS-9T). The same instrument was used to measure heat capacity and electrical transport properties. Simultaneous measurements of longitudinal ($\rho_{xx}$) and Hall resistivity ($\rho_{yx}$ or $\rho_{zx}$) were performed using the conventional four-probe method with a low-frequency excitation current. The electrical transport measurements were carried out on rectangular bar samples, as illustrated in Fig.~1\textbf{c}, where we adopt the crystallographic convention that the $x$, $y$, and $z$ directions correspond to the $[2\bar{1}\bar{1}0]$, $[01\overline{1}0]$, and $[0001]$ axes of the hexagonal unit cell, respectively. In the first configuration, $\rho_{xx}$ and $\rho_{yx}$ were measured with $B \parallel z$ whereas in the second configuration, $\rho_{xx}$ and $\rho_{zx}$ were measured with $B \parallel y$ and for both the cases $I \parallel x$. To eliminate the influence of contact electrode misalignment, the Hall resistivity data were antisymmetrized using the standard expression $\rho_{yx}(B) = [\rho_{yx}(+B) - \rho_{yx}(-B)]/2$.
\par
\textit{DFT calculations.} First-principles calculations were performed within the framework of density functional theory (DFT), as implemented in the Vienna \textit{ab initio} Simulation Package (VASP)~\cite{hafner2008ab}. The exchange-correlation functional was treated using the generalized gradient approximation (GGA)~\cite{perdew} in the Perdew-Burke-Ernzerhof (PBE)~\cite{kresse1999ultrasoft} form. To account for the localized nature of the Mn 3$d$ electrons and their strong electronic correlations, the GGA$+U$ method was employed within the Dudarev approach, with an effective Hubbard parameter $U_{\text{eff}} = U - J = 3.0$ eV ~\cite{dft+U}. A plane-wave energy cutoff of 550 eV was used throughout all calculations. The Brillouin zone was sampled using a Monkhorst-Pack $k$-point mesh of $10 \times 10 \times 6$, centered at the $\Gamma$ point. Structural relaxations were performed until the forces on each atom were less than 0.01 eV/\AA{} and the total energy convergence criterion was set to $10^{-6}$ eV. All properties were calculated using the fully relaxed equilibrium crystal structures.
\par
\textit{Berry curvature and anomalous Hall conductivity.}
The intrinsic AHC was evaluated by explicitly incorporating spin-orbit coupling (SOC) in the self-consistent DFT calculations. The intrinsic Hall conductivity $\sigma_{xy}$ was obtained by integrating the $z$-component of the Berry curvature over all occupied states in the Brillouin zone, as described by\cite{xiao2010berry}:

\begin{equation}
\label{eq:sigma_xy}
\sigma_{xy} = -\frac{e^2}{\hbar} \int \frac{d^3\mathbf{k}}{(2\pi)^3} \sum_n \Omega^z_n(\mathbf{k}) f_n(\mathbf{k}),
\end{equation}

where $f_n(\mathbf{k})$ is the Fermi-Dirac distribution function, and the summation runs over all occupied bands. The Berry curvature $\Omega^z_n(\mathbf{k})$ for the $n$th band is computed using the Kubo-Greenwood-like formula:

\begin{equation}
\label{eq:berry_curvature}
\Omega^z_n(\mathbf{k}) = -2\,\mathrm{Im} \sum_{m \ne n} \frac{ \langle \psi_{n\mathbf{k}} | v_x | \psi_{m\mathbf{k}} \rangle \langle \psi_{m\mathbf{k}} | v_y | \psi_{n\mathbf{k}} \rangle }{ \left[E_m(\mathbf{k}) - E_n(\mathbf{k})\right]^2 },
\end{equation}

where $\psi_{n\mathbf{k}}$ and $E_n(\mathbf{k})$ denote the Bloch wavefunction and eigenvalue of the $n$th band at wavevector $\mathbf{k}$, respectively. The velocity operator is defined as $v_i = \frac{1}{\hbar} \frac{\partial H(\mathbf{k})}{\partial k_i}$, with $i = x, y, z$.

\vspace{0.5em}
\textit{Note added.}~During the preparation of this manuscript, a related work reporting AHE in MgMn$_6$Sn$_6$ was published \cite{ma2025anomalous}.

\section*{ACKNOWLEDGMENTS}

NK acknowledges DST for financial support through Grant Sanction No. CRG/2021/002747 and Max Planck Society for funding under the Max Planck-India partner group project. This research project made use of the instrumentation facility provided by the Technical Research Centre (TRC) at the S.N. Bose National Centre for Basic Sciences, under the Department of Science and Technology (DST), Government of India. KD acknowledges financial support from the DST, India, through a fellowship. We acknowledge access to the single-crystal X-ray facilities at the ISIS Facility for structural characterization. We also acknowledge Subhajit Roychowdhury and Pranav Negi at the Indian Institute of Science Education and Research Bhopal for their support in specific heat measurements. JS and MK acknowledge National Supercomputing Mission (NSM) for providing computing resources of ‘PARAM RUDRA’ at S.N. Bose National Centre for Basic Sciences, which is implemented by C-DAC and supported by the Ministry of Electronics and Information Technology (MeitY) and the DST, Government of India. 

\label{sec:acknowledgments}

\newpage

\providecommand{\noopsort}[1]{}\providecommand{\singleletter}[1]{#1}%

\end{document}